\documentclass[a4paper]{jpconf}
\usepackage{graphicx}
\begin{document}
\title{Ternary cluster decay within the liquid drop model}

\author{G Royer, K Degiorgio, M Dubillot and E Leonard}

\address{Subatech, Universit\'e-IN2P3/CNRS-Ecole des Mines, 44307 Nantes, France}

\ead{royer@subatech.in2p3.fr}

\begin{abstract}
Longitudinal ternary and binary fission barriers of $^{36}$Ar, $^{56}$Ni and $^{252}$Cf nuclei have been determined within
a rotational liquid drop model taking into account the nuclear proximity energy. For the light nuclei the heights
of the ternary fission barriers become competitive with the binary ones at high angular momenta since the maximum 
lies at an outer position and has a much higher moment of inertia.
\end{abstract}

\section{Introduction}  
The observation of events with a missing mass consisting of 2, 3 and 4 $\alpha$ particles in the $^{32}S$+$^{24}Mg$, $^{36}Ar$+$^{24}Mg$ and $^{24}Mg$+$^{12}C$ reactions \cite{zher07,ange06} and the observation of the $\alpha$ and $^{10}Be$ emission in ternary cold neutronless spontaneous fission of $^{252}Cf$ \cite{rama298,dani04} have renewed interested in investigating the ternary fission mode \cite{poen01,deli03,wage04}. Furthermore, ternary $\alpha$ emission has been detected recently during the fission of the rotating compound nucleus $^{236}U$ \cite{gon07}. 

The longitudinal ternary fission has been investigated previously using a family of compact and creviced shapes, or quasi-molecular elongated shapes, leading to the emission of a light fragment between two equal heavier fragments keeping a symmetry plane perpendicular to the fission axis \cite{mign87,roye90,roye92,royer92}. In the present study, the approach is extended to the longitudinal ternary fission with emission
of three different aligned fragments and is applied to several nuclei investigated experimentally. The deformation energy is always calculated within a liquid drop model taking into account the nuclear proximity energy resulting from the strong interaction between the surfaces in regard in the crevices of the one-body strongly deformed configuration or the necks between the separated fragments.

\section{Contributions to the total deformation energy}
\begin{figure}[htbp]
\begin{center}
\includegraphics[trim = 3cm 2cm 1cm 0cm, clip,width=4.64cm,angle=-90]{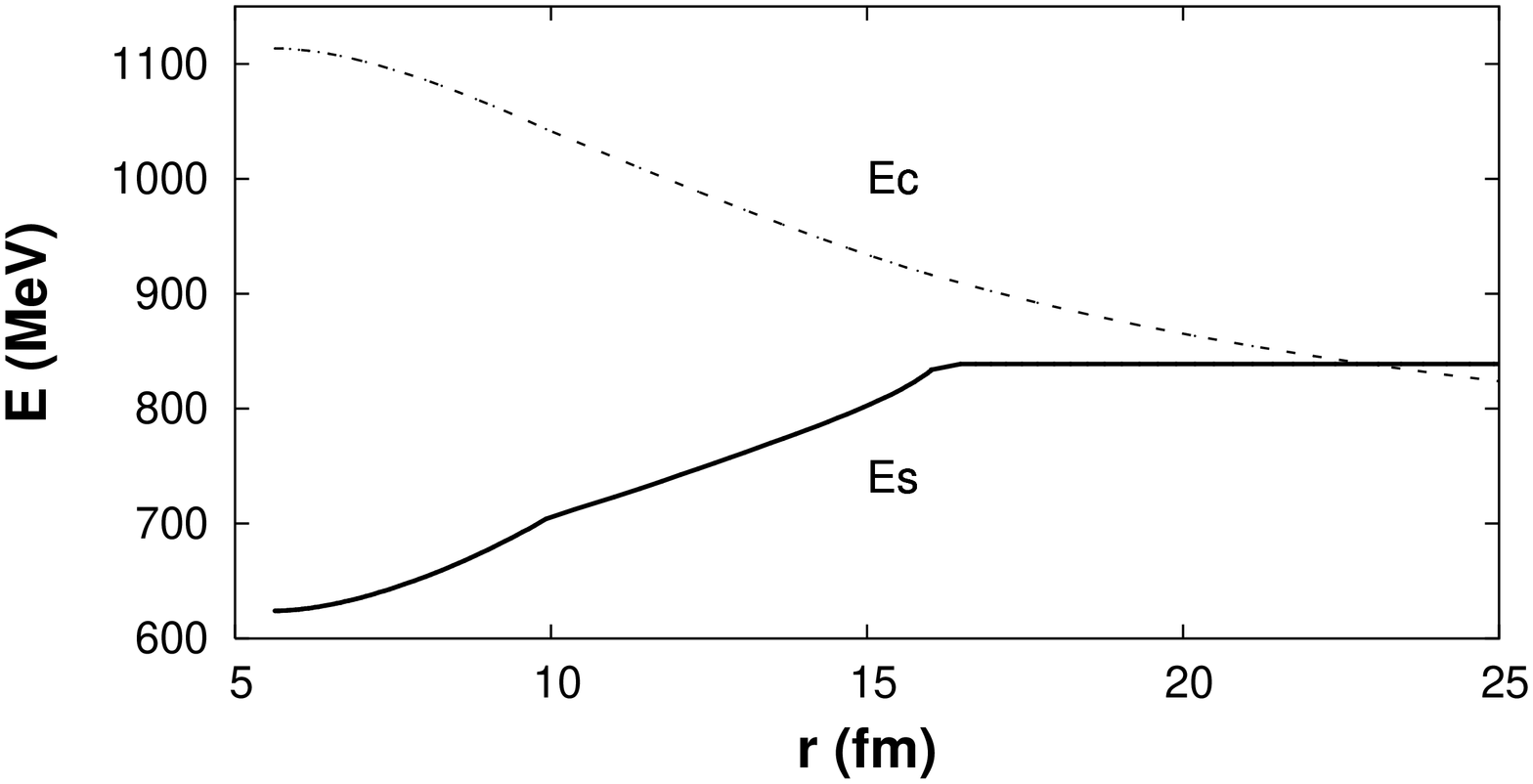}\\[-2.1cm]
\includegraphics[trim = 3cm 1.25cm 1cm 0cm, clip,width=4.48cm,angle=-90]{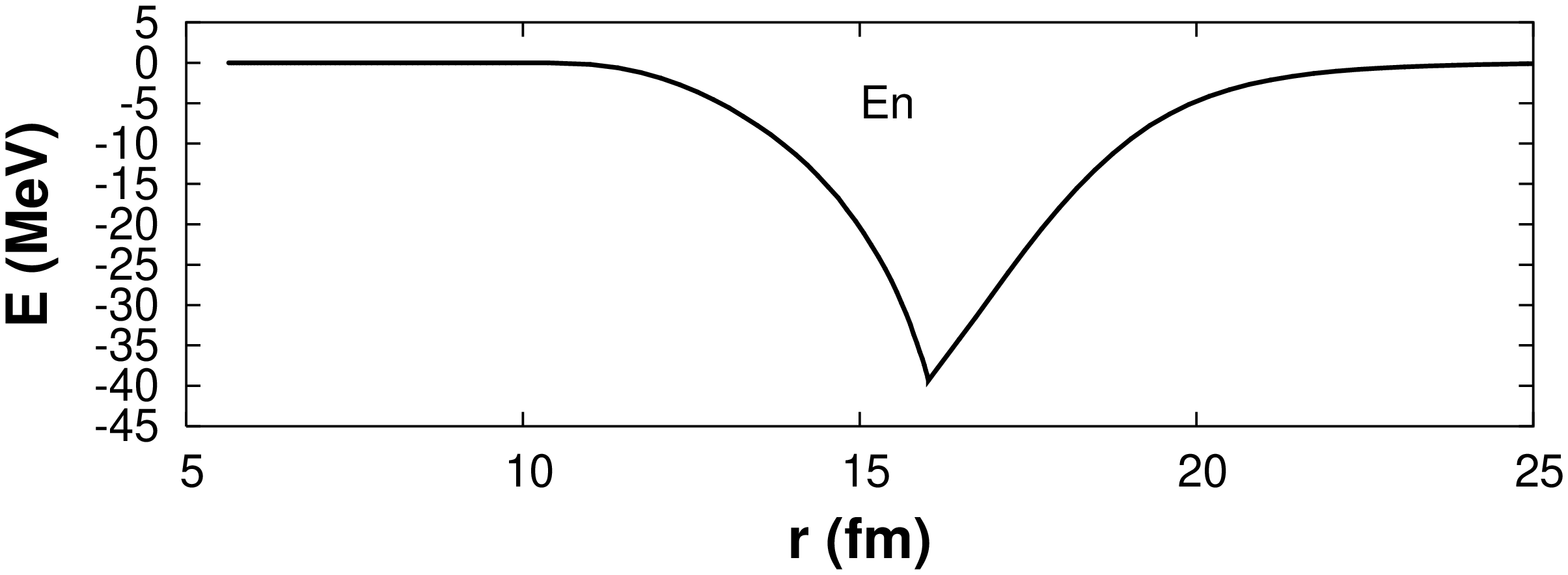}\\[-1.6cm]
\includegraphics[trim = 0cm 0cm 0cm 0cm, clip,width=5.44cm,angle=-90]{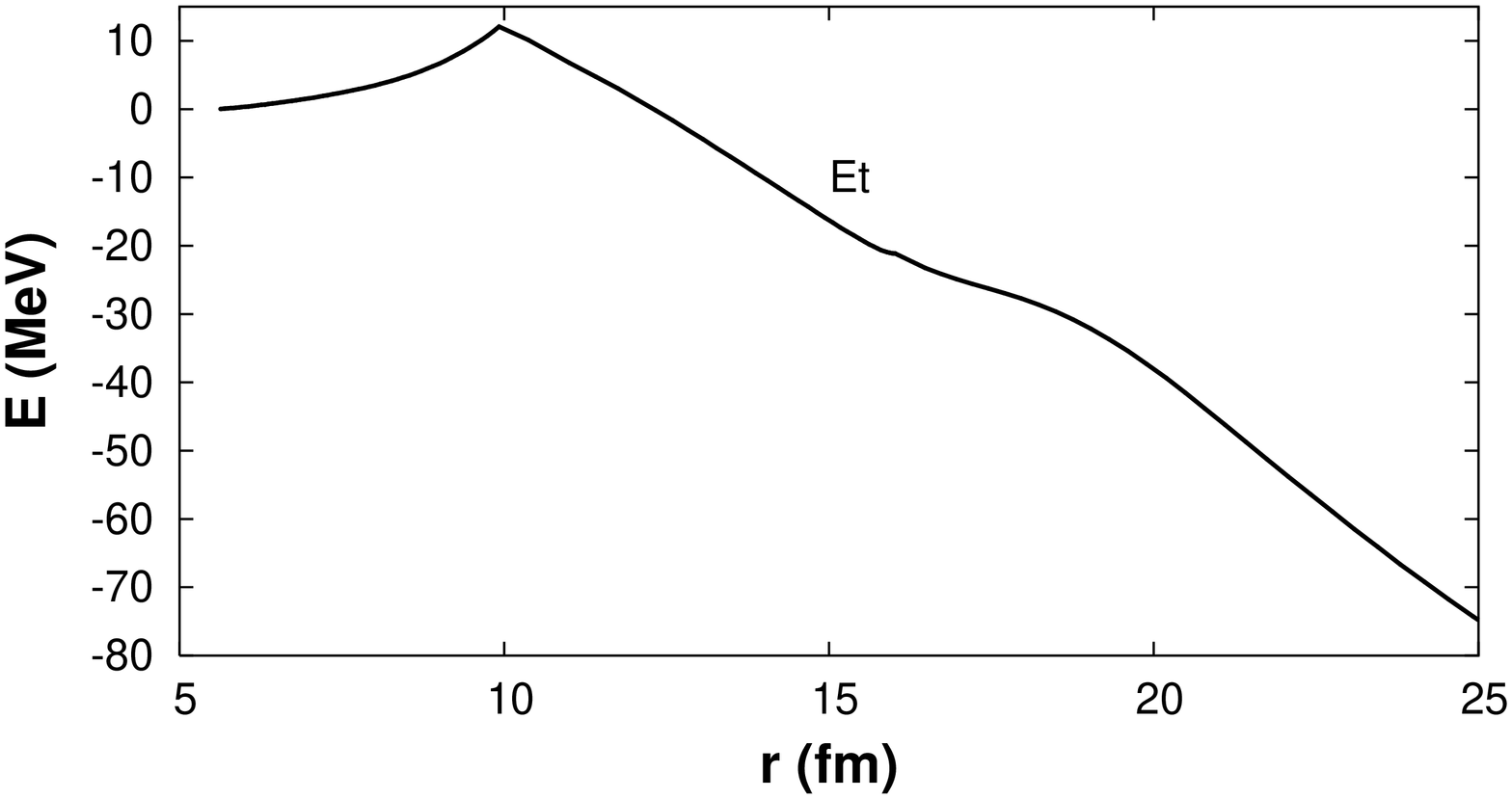}\\
\end{center}
\caption{Contributions of the Coulomb energy $E_c$, surface energy $E_s$ and nuclear proximity energy $E_n$ to the total deformation energy $E_t$ for the $^{121}Ag+^{10}Be+^{121}Ag$ longitudinal ternary decay of $^{252}Cf$.}
\end{figure}

For a deformed nucleus, the macroscopic energy is defined as
\begin{equation}
E=E_{V}+E_{S}+E_{C}+E_{Rot}+E_{Prox}.
\end{equation}
For one-body shapes the volume, surface and Coulomb  energies are defined as :
\begin{equation}
E_{V}=-15.494(1-1.8I^2)A \ {MeV},
\end{equation}
\begin{equation}
E_{S}=17.9439(1-2.6I^2)A^{2/3}(S/4\pi R_0^2) \ {MeV},
\end{equation}
\begin{equation}
E_{C}=0.6e^2(Z^2/R_0) \times 0.5\int
(V(\theta)/V_0)(R(\theta)/R_0)^3 \sin \theta d \theta.
\end{equation}
$S$ is the surface of the one-body deformed nucleus. $V(\theta )$
is the electrostatic potential at the surface and $V_0$ the
surface potential of the sphere.

For three-body shapes, the volume, surface and Coulomb energies are the sum
of the contributions of each fragment and the Coulomb interaction energy is added. 

The rotational energy is determined within the rigid-body ansatz :
\begin{equation}
    E_{Rot}=\frac{\hbar^{2}l(l+1)}{2I_{\bot}}.
\end{equation}
The surface energy results from the effects of the surface tension
forces in a half space. When there are nucleons in regard in a
neck or a gap between separated fragments an
additional proximity energy must be added to take into
account the effects of the nuclear forces between the close
surfaces. This term is essential to describe smoothly the one-body
to two-body transition and to obtain reasonable fusion barrier
and $\alpha$ decay barrier heights. 
\begin{equation}
E_{Prox}(r)=2\gamma \int _{h_{min}} ^{h_{max}} \Phi \left \lbrack
D(r,h)/b\right \rbrack 2 \pi hdh,
\end{equation}
where $h$ is the distance varying from the neck radius or zero to
the height of the neck border. $D$ is the distance between the
surfaces in regard and $b=0.99$~fm the surface width. $\Phi$ is the
proximity function. The surface
parameter $\gamma$ is the geometric mean between the surface
parameters of the two nuclei or fragments. 

The Coulomb, surface and nuclear proximity energies contributing to the ternary fission barrier of  $^{252}Cf$ are displayed in figure 1 for a specific exit channel. The surface and proximity energy slopes change drastically at the contact point between the nascent fragments but the total energy varies smoothly as well as the Coulomb energy. 

\begin{figure}[htbp]
\includegraphics[width=5cm,angle=-90]{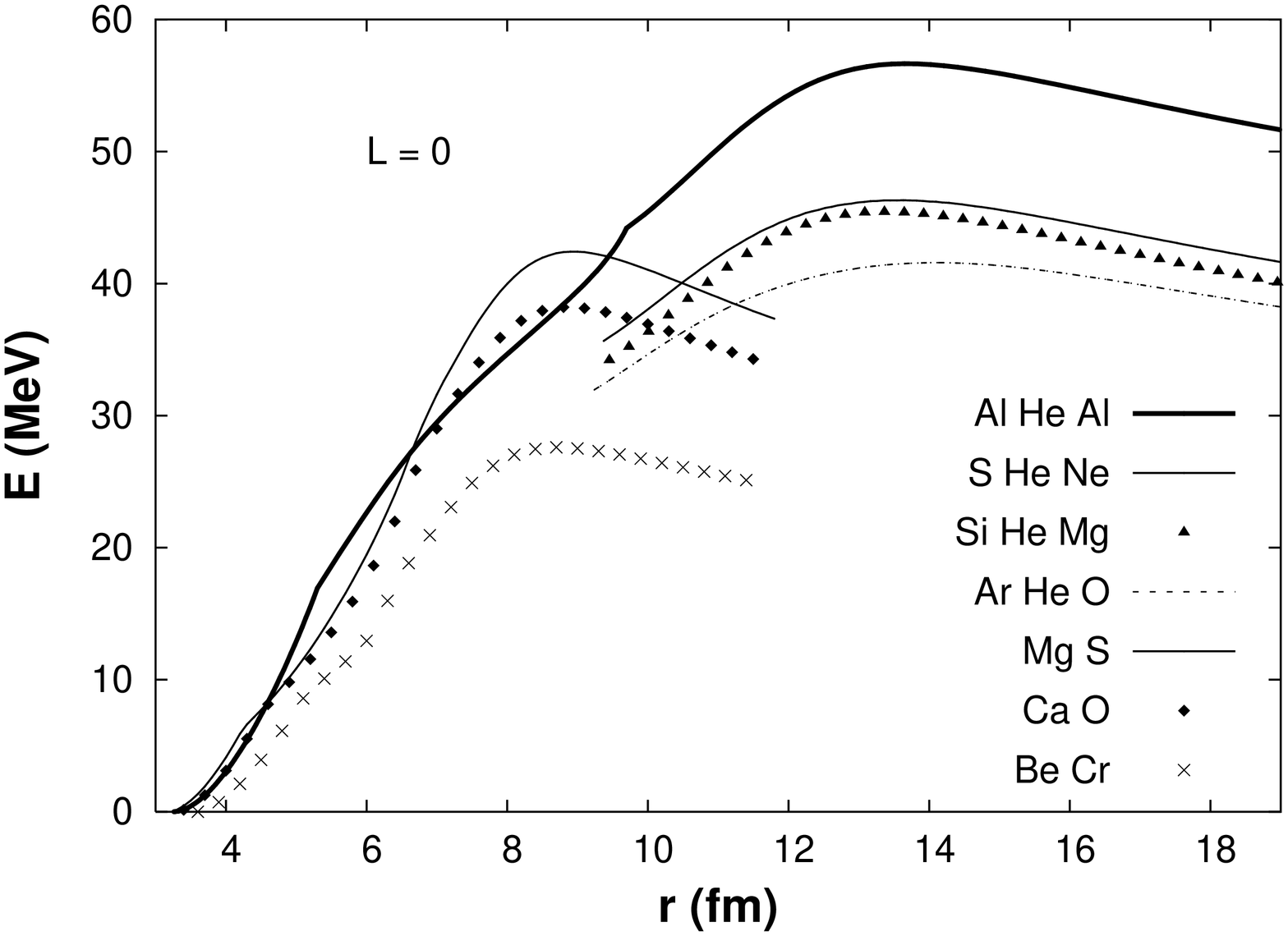}
\includegraphics[width=5cm,angle=-90]{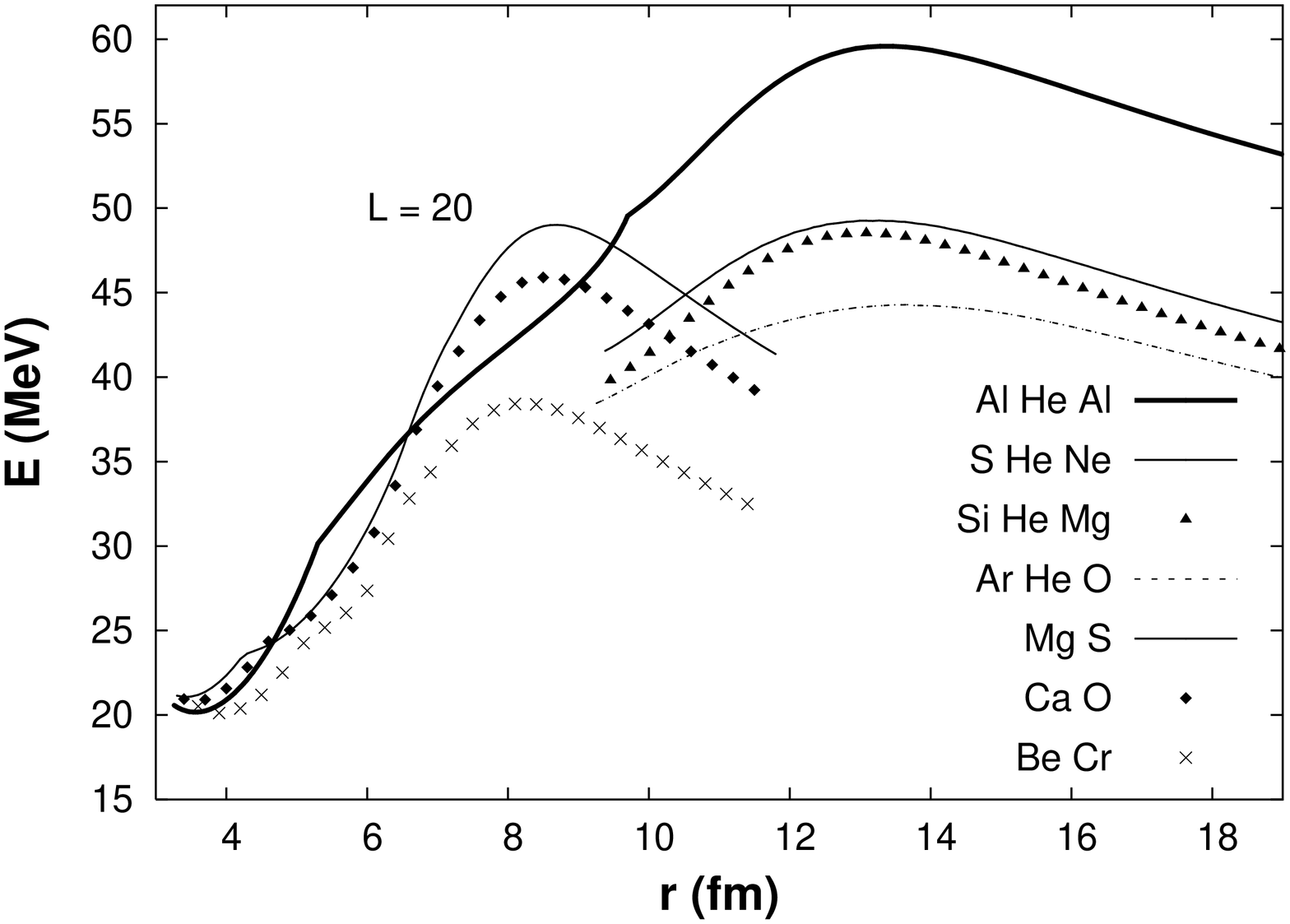}
\centerline{\includegraphics[width=5cm,angle=-90]{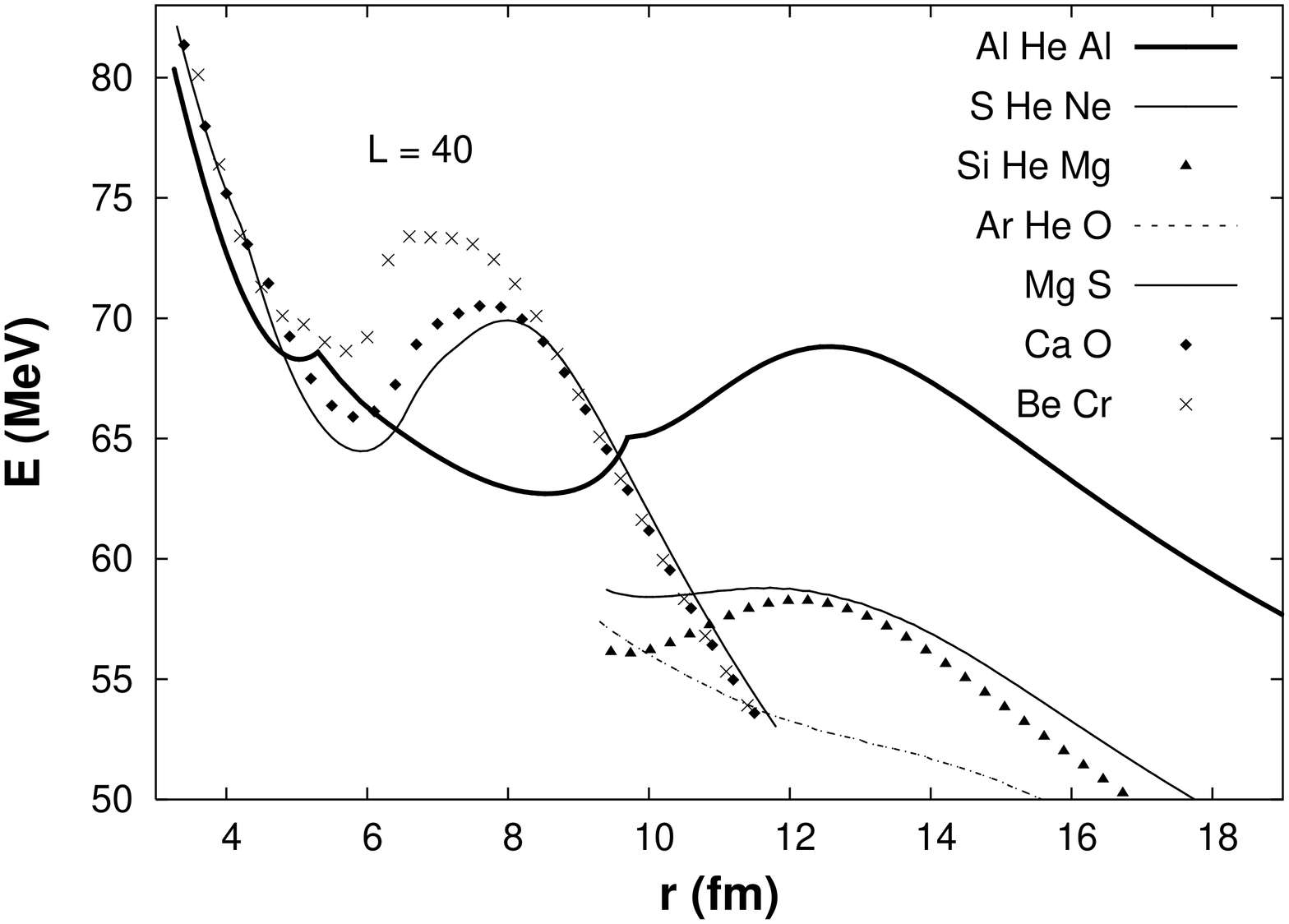}}
\caption{Ternary and binary fission barriers of $^{56}Ni$ at $L=0, 20, 40~\hbar$ leading to $N=Z$ fragments and ternary $^4He$ emission.}
\end{figure}
\section{Ternary and binary potential barriers of $^{56}Ni$}
The potential barriers governing the ternary fission of $^{56}Ni$ leading to the $^{4}He$, $^{8}Be$ or $^{12}C$ emission  
are given in figures 2, 3 and 4 as functions of the angular momentum. The barrier top corresponds always to three or two separated spherical aligned fragments maintained in unstable equilibrium by the balance between the attractive nuclear proximity forces and the repulsive Coulomb forces. The position of the barrier is much more external in the ternary decay path than in the binary one since the elongation is larger. The proximity energy is higher in the ternary fission path since there are two necks. For the ternary $\alpha$ emission and at $L=0$ the ternary fission is not competitive with respect to binary fission. The asymmetric binary fission is favoured. With increasing angular momentum the situation is reversed and the most asymmetric ternary fission may play an important role. The same conclusion can be drawn for the 2 $\alpha$ emission even though the lowering of the ternary fission is weaker. A deep minimum exists in the $^{24}Mg+^{8}Be+^{24}Mg$ exit channel at $L=40~\hbar$ and the transition to an asymmetric ternary decay channel seems possible. The $^{12}C$ emission at $L=40~\hbar$ appears also possible in an asymmetric ternary decay. Then, at the highest angular momenta, the more negative Q value for the ternary fission can be compensated for the smaller value of the rotational energy at the saddle point which occurs at larger distances.       

The same behaviour is observed for the $^{36}Ar$ nucleus.

\begin{figure}[htbp]
\includegraphics[width=5cm,angle=-90]{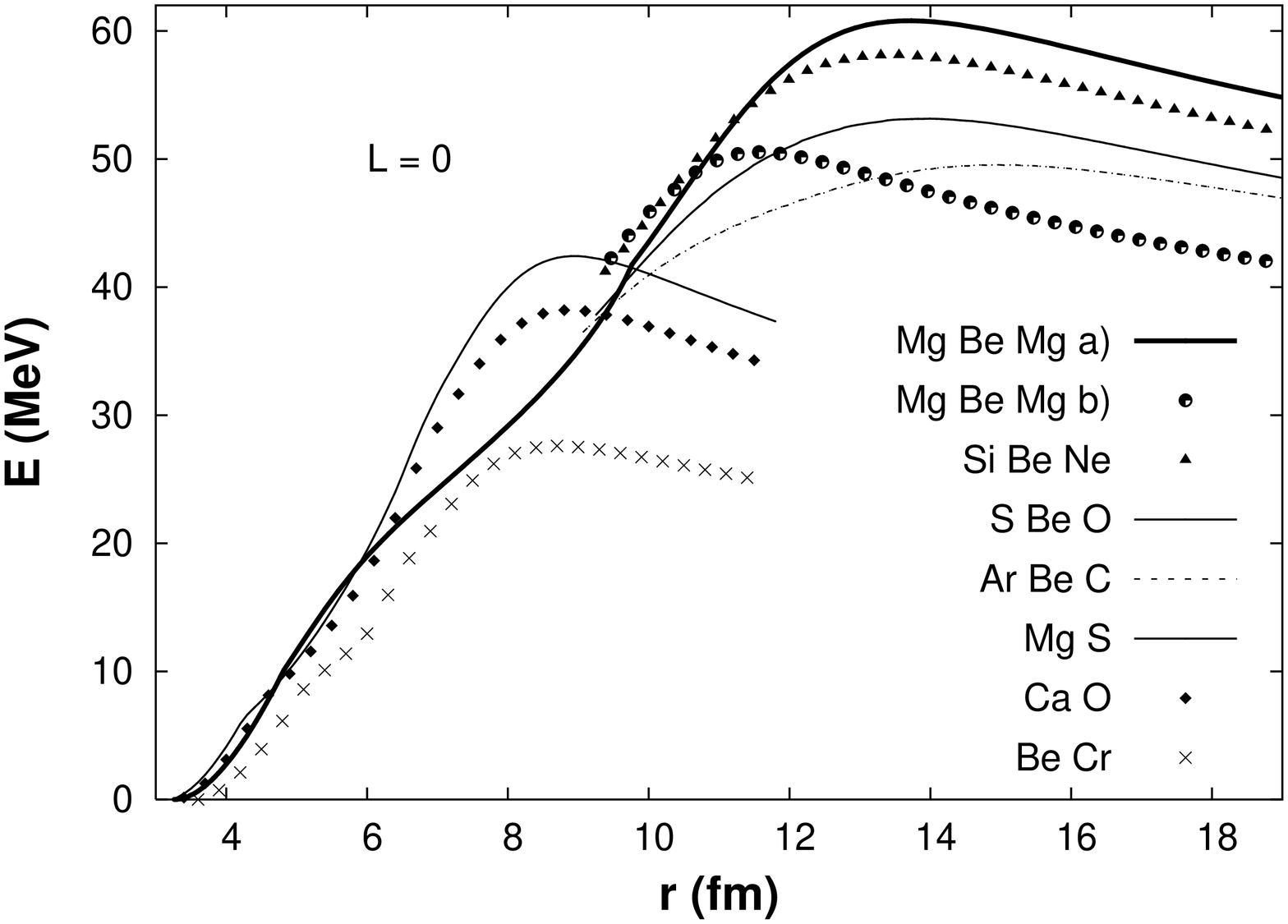}
\includegraphics[width=5cm,angle=-90]{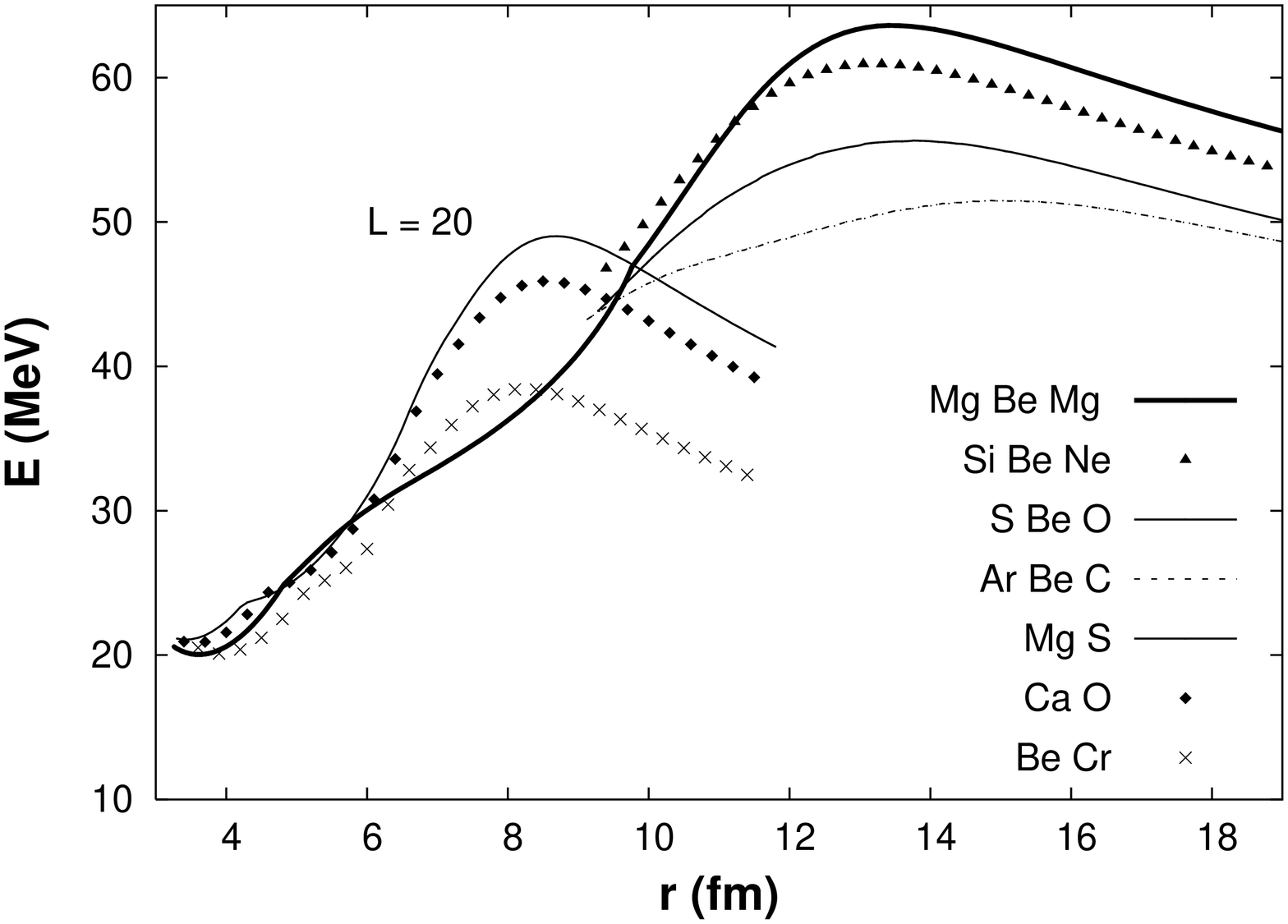}
\centerline{\includegraphics[width=5cm,angle=-90]{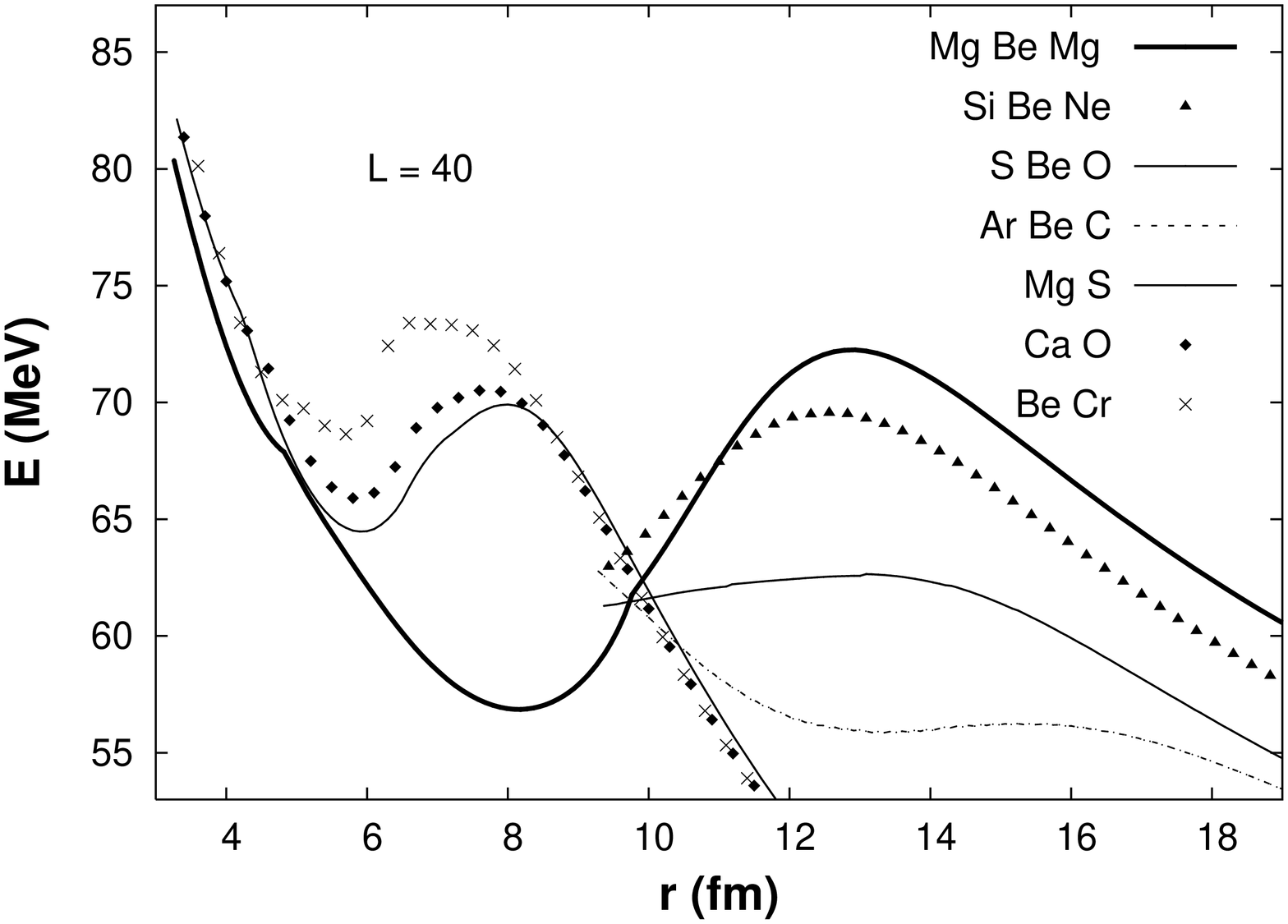}}
\caption{Ternary and binary fission barriers of $^{56}Ni$ at $L=0, 20, 40~\hbar$ leading to $N=Z$ fragments and ternary $^8Be$ emission.}
\end{figure}

\begin{figure}[htbp]
\includegraphics[width=5cm,angle=-90]{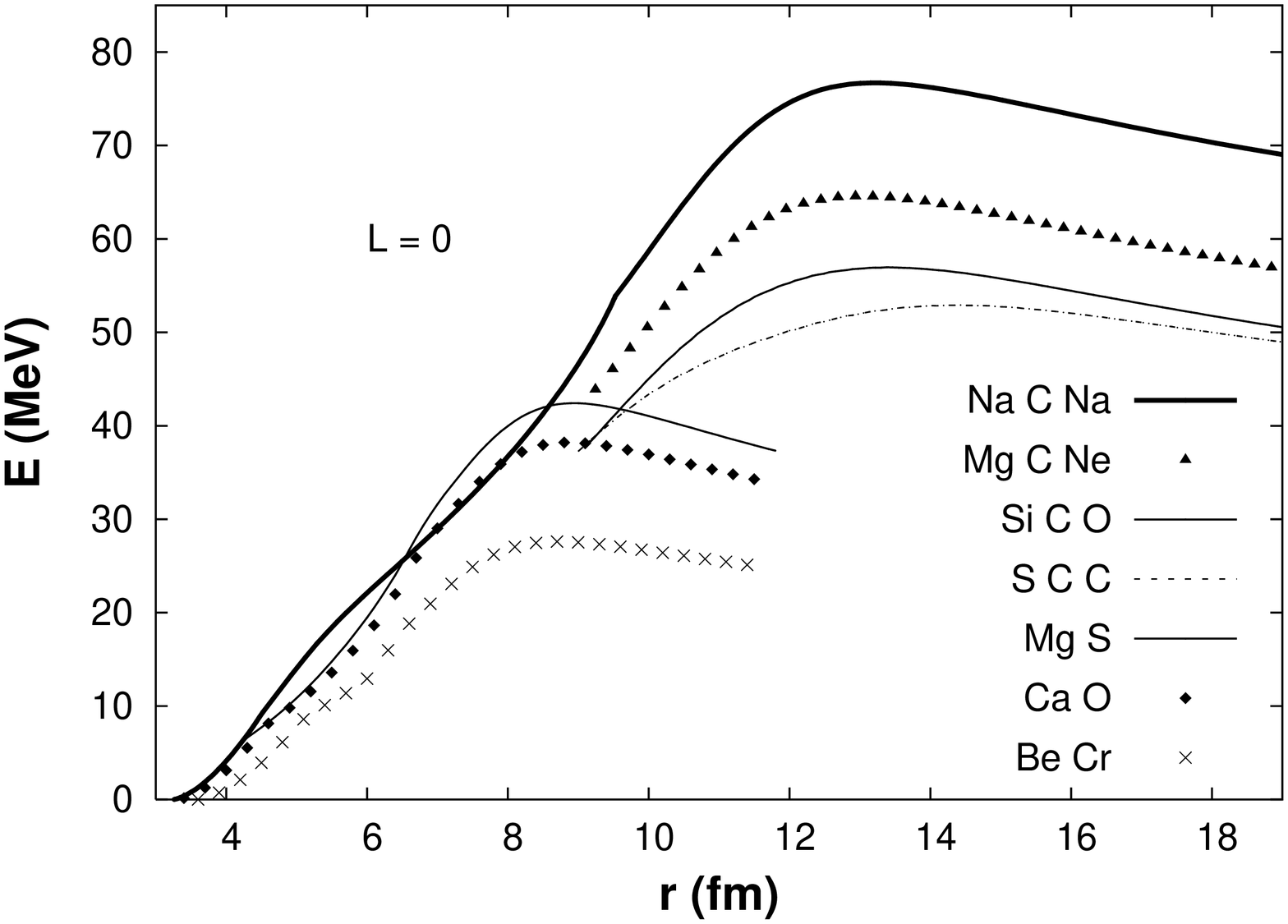}
\includegraphics[width=5cm,angle=-90]{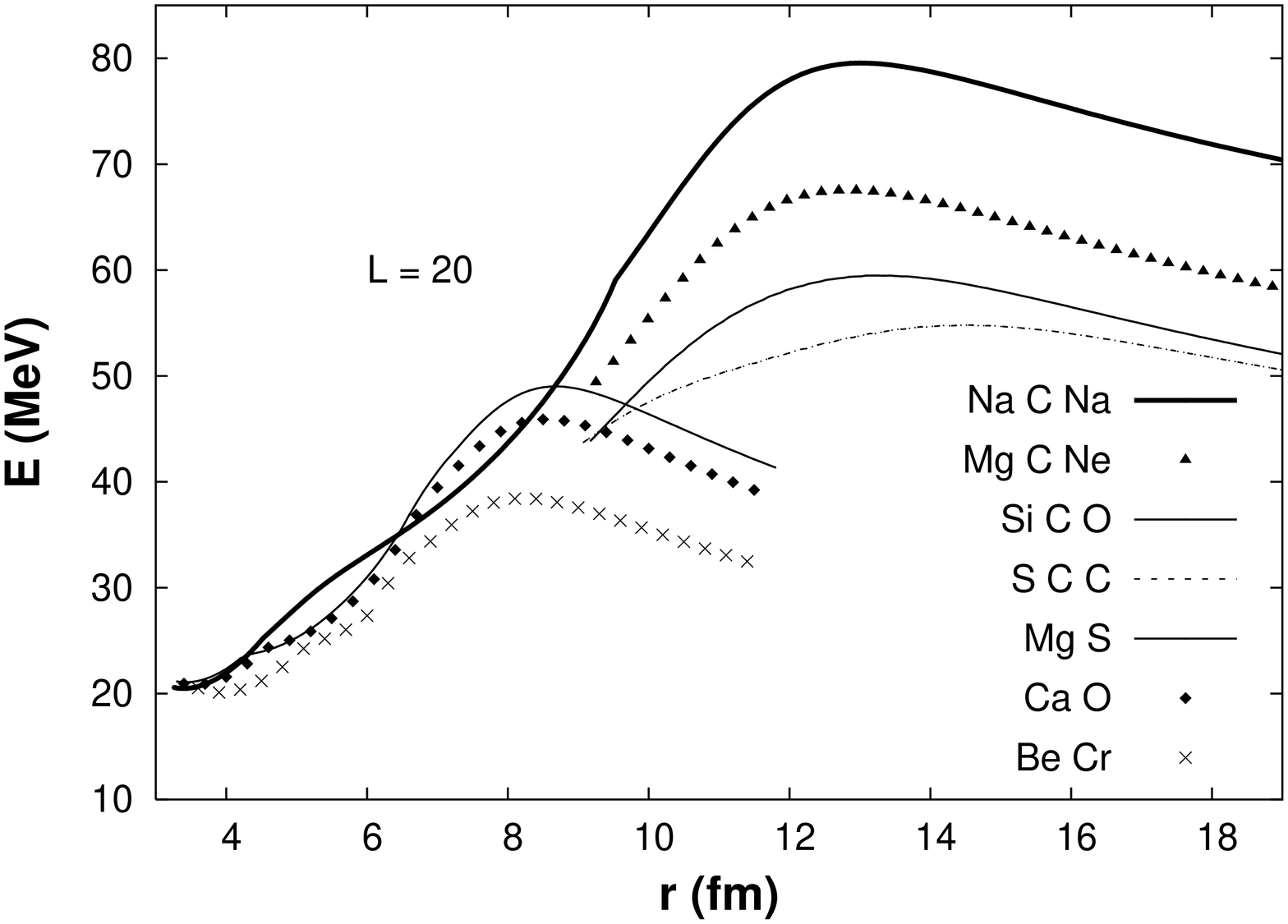}
\centerline{\includegraphics[width=5cm,angle=-90]{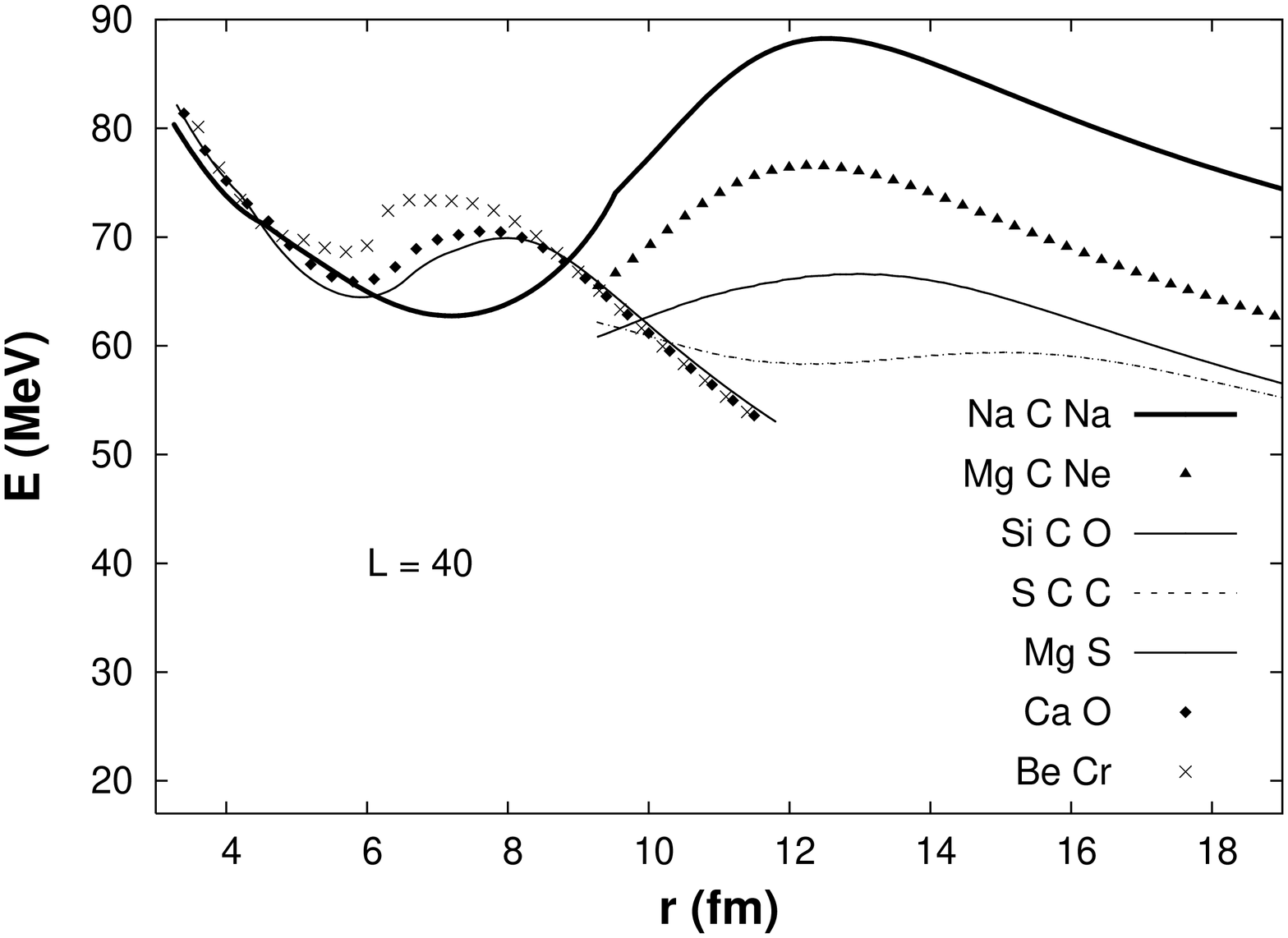}}
\caption{Ternary and binary fission barriers of $^{56}Ni$ at $L=0, 20, 40~\hbar$ leading to $N=Z$ fragments and ternary $^{12}C$ emission.}
\end{figure}

\section{Barrier characteristics}
The fission barrier heights of the $^{56}Ni$ nucleus and the moment of inertia at the top of the barrier are given in the table 1. The moment of inertia in the ternary decay is roughly twice larger than in the binary decay channel. The moment of inertia increases with the symmetry of the reaction. The moment of inertia diminishes when the angular momenta increases since the barrier top moves slightly to an inner position.    
\begin{table}
\caption{\label{Table1}Binary and ternary fission barrier height (in MeV) and moment of inertia (in $\hbar^2$/MeV) at the maximum for $^{56}Ni$.}
\begin{center}
\lineup
\item[]\begin{tabular}{llll}
\br
Reaction &  L &  Emax & Moment \\
\mr
$^{8}Be$ + $^{48}Cr$ & \00& 30.6 & 20.1 \\
$^{16}O$ + $^{40}Ca$ & \00  & 38.2 & 28.1 \\
$^{24}Mg$ + $^{32}S$ & \00  & 42.4 & 32.7 \\
$^{16}O$ + $^{4}He$ + $^{36}Ar$ & \00  & 41.6 & 70.2 \\
$^{20}Ne$ + $^{4}He$ + $^{32}S$ & \00 & 46.3 & 70.1 \\
$^{24}Mg$ + $^{4}He$ + $^{28}Si$ & \00  & 45.4 & 69.3 \\
$^{12}C$ + $^{8}Be$ + $^{36}Ar$ & \00  & 49.6 & 79.4 \\
$^{16}O$ + $^{8}Be$ + $^{32}S$ & \00  & 53.0 & 75.5 \\
$^{20}Ne$ + $^{8}Be$ + $^{28}Si$ & \00  & 58.1 & 73.7 \\
$^{24}Mg$ + $^{8}Be$ + $^{24}Mg$ & \00 & 59.7 & 71.6 \\
$^{12}C$ + $^{12}C$ + $^{32}S$ & \00  & 52.9& 80.2 \\
$^{16}O$ + $^{12}C$ + $^{28}Si$ & \00& 57.0 & 73.2 \\
$^{20}Ne$ + $^{12}C$ + $^{24}Mg$ & \00  & 64.6 & 70.7 \\
$^{8}Be$ + $^{48}Cr$ & 20  & 41.4 & 18.8 \\
$^{16}O$ + $^{40}Ca$ & 20  & 45.9 & 26.5 \\
$^{24}Mg$ + $^{32}S$ & 20 & 42.0 & 31.2 \\
$^{16}O$ + $^{4}He$ + $^{36}Ar$ & 20  & 44.3 & 66.2 \\
$^{20}Ne$ + $^{4}He$ + $^{32}S$ & 20 & 49.3 & 66.6 \\
$^{24}Mg$ + $^{4}He$ + $^{28}Si$ & 20 & 48.5 & 66.7 \\
$^{12}C$ + $^{8}Be$ + $^{36}Ar$ & 20 & 51.5 & 78.1 \\
$^{16}O$ + $^{8}Be$ + $^{32}S$ & 20 & 55.6 & 73.0 \\
$^{20}Ne$ + $^{8}Be$ + $^{28}Si$ & 20 & 61.0 & 70.3 \\
$^{24}Mg$ + $^{8}Be$ + $^{24}Mg$ & 20 & 62.7 & 69.9 \\
$^{12}C$ + $^{12}C$ + $^{32}S$ & 20 & 54.8 & 81.6 \\
$^{16}O$ + $^{12}C$ + $^{28}Si$ & 20 & 59.5 & 73.3 \\
$^{20}Ne$ + $^{12}C$ + $^{24}Mg$ & 20 & 67.5 & 69.1 \\
$^{8}Be$ + $^{48}Cr$ & 40 & 76.4 & 16.0 \\
$^{16}O$ + $^{40}Ca$ & 40 & 70.5 & 23.1 \\
$^{24}Mg$ + $^{32}S$ & 40& 69.9 & 27.4 \\
$^{24}Mg$ + $^{4}He$ + $^{28}Si$ & 40  & 58.3 & 58.3 \\
$^{16}O$ + $^{8}Be$ + $^{32}S$ & 40& 62.6 & 66.0 \\
$^{20}Ne$ + $^{8}Be$ + $^{28}Si$ & 40 & 69.6 & 64.5 \\
$^{24}Mg$ + $^{8}Be$ + $^{24}Mg$ & 40 & 71.9 & 63.1 \\
$^{16}O$ + $^{12}C$ + $^{28}Si$ & 40 & 66.6 & 68.0 \\
$^{20}Ne$ + $^{12}C$ + $^{24}Mg$ & 40 & 76.5 & 62.7 \\
\br
\end{tabular}
\end{center}
\end{table}

\section{Conclusion}
The longitudinal asymmetric ternary fission has been investigated within a generalized liquid drop model taking into account the influence of the proximity forces. For the light nuclei the heights
of the ternary fission barriers become competitive with the binary ones at high angular momenta since the maximum 
lies at an outer position and has a much higher moment of inertia. Then the probability of ternary fission events which is negligible at low angular momentum strongly increases at the highest angular momenta. This might explain the possible observation of events with a missing mass of 2, 3 and 4 $\alpha$ particles in the $^{32}S$+$^{24}Mg$ and $^{36}Ar$+$^{24}Mg$   reactions.


\section{References}
\begin{thebibliography}{15}
\bibitem{zher07}
Zherebchevsky V, v. Oertzen W, Kamanin D, Gebauer B, Thummerer S, Schulz C and Royer G 2007 {\it Phys. Lett. B} {\bf 646} 12
\bibitem{ange06}
S\`anchez i Zafra A 2006 thesis (IPHC 06-010, Strasbourg)
\bibitem{rama298}
Ramayya A V {\it et al} 1998 {\it Phys. Rev. Lett. } {\bf 81} 947
\bibitem{dani04}
Daniel A V {\it et al} 2004 {\it Phys. Rev. C} {\bf 69} 041305(R)
\bibitem{poen01}
Poenaru D N, Greiner W, Hamilton J H and Ramayya A V 2001 {\it J. Phys. G: Nucl. Part. Phys.} {\bf 27} L19
\bibitem{deli03}
Delion D S, Sandulescu A and Greiner W 2003 {\it J. Phys. G: Nucl. Part. Phys.} {\bf 29} 317
\bibitem{wage04}
Wagemans C, Heyse J, Janssens P, Serot O and Geltenbort P 2004 {\it Nucl. Phys. A} {\bf 742} 291
\bibitem{gon07}
Goennenwein F {\it et al} 2007 {\it Phys. Lett. B} {\bf 652} 13
\bibitem{mign87}
 Mignen J and Royer G 1987 {\it J. Phys. G: Nucl. Phys.} {\bf 13} 987
\bibitem{roye90}
 Mignen J and Royer G 1990 {\it J. Phys. G: Nucl. Part. Phys.} {\bf 16} L227
\bibitem{roye92}
 Royer G and Mignen J 1992 {\it J. Phys. G: Nucl. Part. Phys.} {\bf 18} 1781
\bibitem{royer92}
 Royer G, Haddad F and Mignen J 1992 {\it J. Phys. G: Nucl. Part. Phys.} {\bf 18} 2015
\end{thebibliography}
\end{document}